\documentclass[useAMS,usenatbib]{mn2e}
\usepackage[applemac]{inputenc}
\usepackage[pdftex]{color, graphicx}
\usepackage{subfigure}
\usepackage{amsmath}
\usepackage{amssymb}

\def\aj{AJ}                 

\title[Phase-space structure and populations in NGC~ 2264]{Phase-space structures and stellar populations in the star-forming region NGC~ 2264}
\author[Marta Gonz\'alez \& Emilio J. Alfaro]{Marta Gonz\'alez\thanks{E-mail: marta@iaa.es}  and Emilio J. Alfaro
 \\
Instituto de Astrof\'{\i}sica de Andaluc\'{\i}a, Glorieta de la Astronom\'ia s/n, Granada E-18008, Spain}
\begin{document}

\date{Accepted year month day. Received year month day; in original form year month day }

\pagerange{\pageref{}--\pageref{}} \pubyear{2016}

\maketitle

\label{firstpage}

\begin{abstract}

In this work we analyse the structure of a subspace of the phase space of the star-forming region NGC~ 2264 using the Spectrum of Kinematic Groupings (SKG). We show that the SKG can be used to process a collection of star data to find substructure at different scales. We have found structure associated with the NGC~ 2264 region and also with the background area. 
In the NGC~ 2264 region, a hierarchical analysis shows substructure compatible with that found in previous specific studies of the area but with an objective, compact methodology that allows us to homogeneously compare the structure of different clusters and star-forming regions. Moreover, this structure is compatible with the different ages of the main NGC~ 2264 star-forming populations. 
The structure found in the field can be roughly associated with giant stars far in the background, dynamically decoupled from NGC~ 2264, which could be related either with the Outer Arm or Monoceros Ring. 
The results in this paper confirm the relationship between structure in the RV phase-space subspace and different kinds of populations, defined by other variables not necessarily analysed with the SKG, such as age or distance, showing the importance of detecting phase-space substructure in order to trace stellar populations in the broadest sense of the word. 

\end{abstract}

\begin{keywords}
stellar clusters -- star-forming regions -- radial velocity -- kinematic structures -- minimum spanning tree -- NGC~ 2264.
\end{keywords}

\section{Introduction}
The study of stellar-forming regions (SFR) and young clusters is key for a complete understanding of cloud collapse and for evaluating star-formation mechanisms. One of the main aims is the search for patterns in the phase-space (in the classical dynamical sense of the term) and its subsequent temporal evolution.

The spatial part of the phase space has been widely studied through a variety of studies and statistical tools \citep[among others]{MST04, SOBA04, Kumar07, Wang08, Schmeja308, Bastian09, Sanchez09, Hetem15}. However, only a few clusters have been studied considering kinematic data as well \citep{Furesz06, Furesz08, Rob14, Sacco15,Tobin15, Belen16}, and all of them used customized exploratory hand-made analyses.
 This lack of knowledge of the structure associated with the kinematic dimensions of the phase-space is due both to the scarcity of accurate and complete kinematic data and also to the absence of reliable statistical tools specifically designed for this purpose. 

In our previous work, \citet{Alf16}, we addressed this issue and presented a statistical tool to analyse the kinematic structure of a dataset, which we call Spectrum of Kinematic Groupings or SKG. We showed the capabilities of this tool for distinguishing structure associated with radial velocity (RV) using a set of test cases with a wide range of configurations. The SKG can be easily implemented in any pipeline developed to mine large databases and leads to a quantitative description of the kinematic pattern allowing a comparative analysis between different clusters, environments and datasets in a homogeneous way. This makes the SKG a suitable procedure for analysing the amount of data expected either from ground-based projects such as APOGEE \citep{APOGEE}, LAMOST \citep{LAMOST}, GES \citep{GES}, and WEAVE \citep{WEAVE}, or from the Gaia space mission \citep{Gaia}. 

In this work we will show the potential of this method for analysing a set of real data from the star-forming region NGC~ 2264. We have chosen this area for two main reasons: first of all, it is a particularly interesting region of the sky where the existence of a spatial pattern of RV has previously been detected \citep{Furesz06,Tobin15}, and at the same time it is one of the regions with most active star formation, containing OB stars, PMS stars, ionized, neutral and molecular gas interacting in the complex process of forming stars.

NGC~ 2264 is a very interesting region, located in the Monoceros OB1 association, in the third Galactic quadrant. This is one of the least obscured regions in the Galactic plane \citep{Oliver96}. Its line of sight may be crossing different Galactic features, such as the Local arm, the Perseus arm, the Outer Arm and the Monoceros Ring. NGC~ 2264 is relatively near, at a distance between 750 and 900 pc \citep{Walker56, SBL97,  Flacco99, Dahm08, Turner12}. \citet{Walker56} discovered for the first time a well-populated sequence of pre-main-sequence stars in this region, in accordance with the young age of its members, between 1 and 6 Myr \citep{SBL97, Flacco99,Turner12} and other hints indicating that star formation has occurred in different regions of the molecular cloud over the last several Myr \citep{Dahm08}. The large amount of molecular gas remaining in the various cloud cores indicates that star formation may continue in the region for several Myr. Despite the gas, NGC~ 2264 shows low reddening \citep{Walker56, Flacco99, Buckle312}, which makes the study of its members easier. Based on optical photometric data of the region, \citet{Sung08} identified two dense SFRs (S MON and CONE in the northern and southern parts of the area, respectively), and a low density halo surrounding them. Later work by \citet{Sungetal09} further refined the CONE SFR, finding two substructures, the Spokes cluster and the Cone Core. 

All these reasons have made NGC~ 2264 a traditional laboratory for studying star formation, and there is a long list of works on this object, focusing on different topics, such as the properties of the gas \citep{Buckle312, Teixeira12}, stellar content \citep{Walker56, Flacco99, Turner12, Zwintz14}, and their interaction and feedback in the cluster region \citep[see][and references therein]{Furesz06, Tobin15}.

In this work we will focus on the results obtained by \citet{Furesz06} and \citet{Tobin15}, which showed a well-structured spatial pattern associated with the radial velocity (RV). \citet{Furesz06} performed a careful and individualized analysis of the data, finding a north-south gradient in RV and spatial substructure associated with RV consistent with the structure of the molecular gas in the region. \citet{Tobin15} expanded the sample from \citet{Furesz06} to perform a new analysis, and found a new population of stars, in this case systematically blueshifted from the molecular gas. 
We will use the complete sample from \citet{Tobin15} to check the quality of the results obtained with the SKG, comparing them with the results from their careful, individualized analysis. 

The paper is divided into four sections, the first being this introduction. The description of the procedure is shown in section 2, and its application to the star-forming region NGC~ 2264 and the main conclusions of the study are presented in sections 3 and 4, respectively.

\section{Method}

The search for phase-space structure in stellar systems requires specific tools that respond to different concepts of what a stellar grouping is. 
Here we consider the existence of a clumpy velocity pattern where there are velocity ranges (channels) whose spatial distribution is more concentrated than that of the whole kinematic interval. In this section we briefly describe the method, but refer to \citep{Alf16} for a thorough description of the foundation and procedure. 

The essence of the method is to examine a group of stars associated with a radial velocity channel, and calculate its \textit{kinematic index} $\tilde \Lambda$, which compares its concentration with that of a group of the same size, representative of the whole sample. To measure the concentration of a group of stars we will use the median edge length of its Euclidean minimum spanning tree (henceforth MST). 

\subsection{Detecting the kinematic groupings}

The procedure designed to search for kinematic segregation starts from a sample of stars in a region of the sky, including both potential cluster members and field stars, preferably with accurate spatial coordinates and precise radial velocity data.

First we sort the data by radial velocity values and separate the sample into bins of size $b$ with an incremental step $s$, so we scan the whole RV domain. We then obtain the MST of each bin and calculate its median edge length. We also calculate the median edge length of the MST of 500 groups of size $b$ randomly extracted from the whole sample, and consider the mean of the 500 medians as the size of a group of $b$ elements representative of the whole sample. 

We estimate the  kinematic index associated with each velocity channel, $RV_{j}$, $\tilde{\Lambda}(RV_{j})$, as

\begin{eqnarray}
	\tilde{\Lambda} (RV_{j}) = 
	\frac {\overline{\tilde{l}^{500}_{Rand}}}   {\tilde{l}_{j}}
	\label{eq2}
\end{eqnarray}
where $ {\tilde{l}_{j}}$ the median edge length of the MST of the bin $j$, and ${\overline{\tilde{l}^{500}_{Rand}}}$ is the mean of the median edge length of the 500 stochastic groups of size $b$. When $\tilde{\Lambda(RV_{j})}>1$ then the bin $j$ is more concentrated than the representative group. 

 A plot of $\tilde{\Lambda} (RV_{j})$ against the median of the radial velocity $\widetilde{RV}$ in the $RV_{j}$ interval for all j, constitutes the \textit{spectrum of kinematic groupings, or SKG,} which provides an interesting and useful tool for the exploratory analysis of the stellar-system phase space. 
 
 Those $\tilde{\Lambda} (RV_{j})$ that verify the inequality 
\begin{eqnarray}
	\frac {\overline{\tilde{l}^{500}_{Rand}}}   {\tilde{l}_{j}} - 
	\frac {2\times\tilde{\sigma}^{500}_{Rand}}  {\tilde{l}_{j}}  \equiv \tilde{\Lambda} (RV_{j}) -2\times \sigma_{\tilde{\Lambda} (RV_{j})} > 1
	\label{eq3}
\end{eqnarray}
mark the radial velocity channels $RV_{j}$ where kinematic structure or segregation has been detected. Eq. \ref{eq3} is a conservative criterion where $\tilde{\Lambda} (RV_{j})$ is larger than unity with confidence level greater than 95\%. 

After applying this method, we get eight variables associated with each RV bin: four of them correspond to its location in the cluster and its spatial extent; another two are the radial velocity median and dispersion; and the last two are the kinematic index and its uncertainty. 

\section{Application to NGC~ 2264}

We now apply the method described in the previous section to the case of NGC~ 2264.
Our sample consists of the (RA, DEC, RV) data from \citet{Tobin15}, downloaded from the VizieR catalogue \citep{Vizier}, for 995 stars (of which 407 are very likely members according to the authors) in the field of NGC~ 2264 with a very broad range of RV. The stars are mainly distributed along a north-south strip with two apparent concentrations or lobes, consistent with the structure described in \citet{Sung08}. The S MON SFR would correspond to the northern lobe and the CONE SFR to the southern. There is also a medium density structure in the central western area of the region. 

The first step was changing the coordinates to angular distances, $\Delta$RA and $\Delta$DEC, to a centroid of the area, in this case the average position in (RA, DEC)=(100.189, 9.691). This was done so the position coordinates used are more coherent with the Euclidean distances we use in the MST calculations. 

\subsection{Raw analysis}


\begin{figure*}
	\centering
	\includegraphics[width=\textwidth]{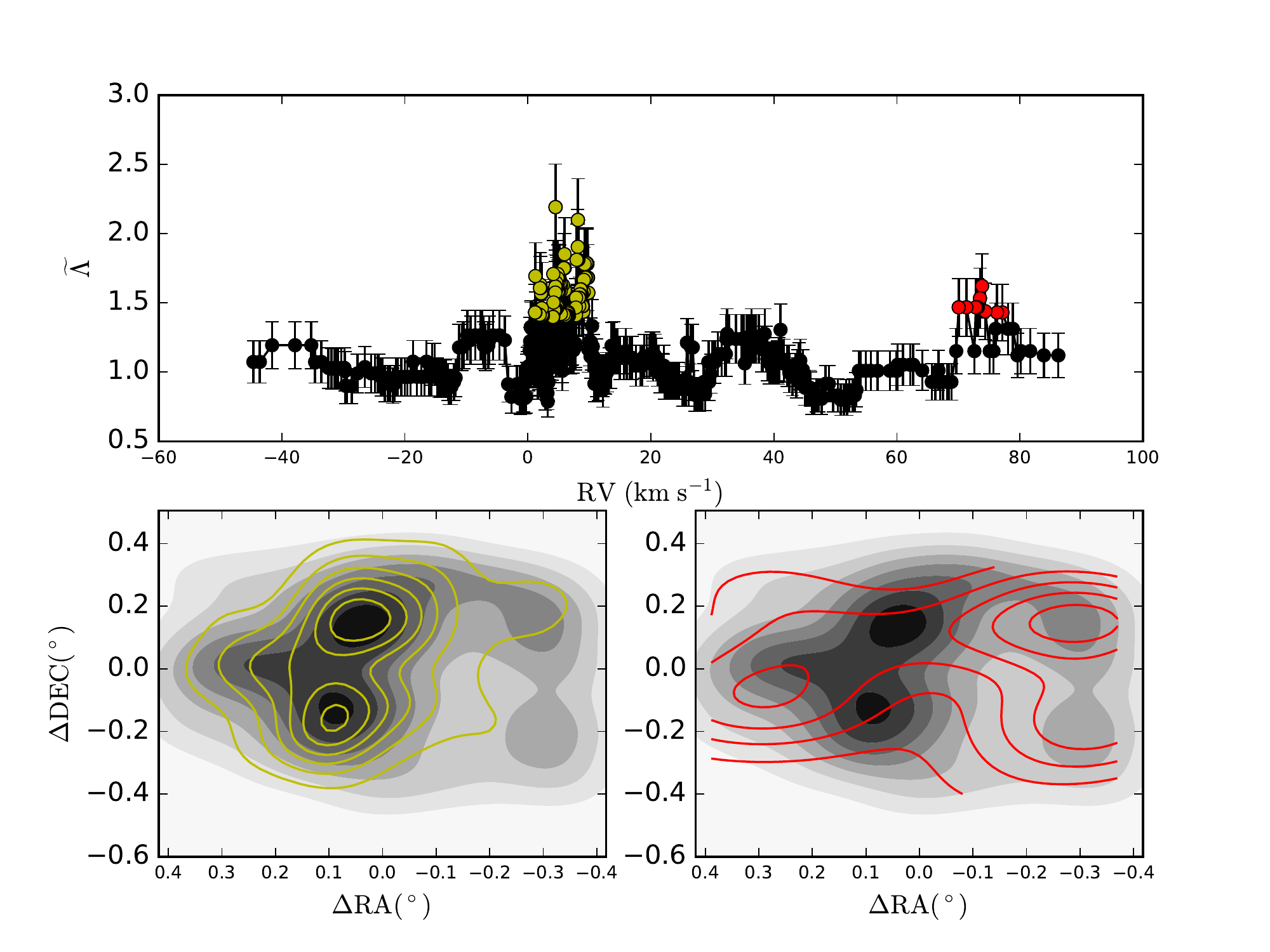}
		\caption{Three-panel summary mosaic of the analysis of the whole sample with bin size $b=26$ and step size $s=1$. Upper left panel: Spectrum of kinematic groupings, as described in the text. The coloured circles mark the segregated velocity intervals according to the inequality defined by equation \ref{eq3}.  Yellow circles represent group A, with median RV 4.98 km/s, and red circles group B, with median RV 73.69 km/s. 
		Lower panels: Density contours of the segregated groups in colour over the greyscale density map of the sample. The colour code is the same as in the upper panel.}
		\label{fig1}
\end{figure*}


We started by feeding our algorithm with all the spatial and RV data of all the stars independently of the RV measurement quality or membership probability, to check the performance of the method in a realistic situation of contaminated, possibly biased samples with error. In addition, we set the bin size to $b=26$, the nearest integer to the square root of the number of stars to analyse, and step $s=1$, as would be done in an automatic pipeline implementation. 

\begin{table}
\caption{Characteristics of selected groups A and B, and the separated components of A: A1, A2, and A3. From left to right: Number of stars, mean radial velocity, median radial velocity and velocity dispersion}
\begin{center}
\begin{tabular}{lcccc}
\hline
Group& N & $\overline{\rm{RV}}$ (km/s) & $\widetilde{\rm{RV}}$ (km/s) & $\sigma_{\rm{RV}}$ (km/s)\\
\hline
A & 279 & 5.14& 4.98&2.59\\
B & 39  & 75.57&73.69&9.51\\
\hline
A1 & 206 & 6.32 & 5.83 & 1.09\\
A2 & 42  & 2.23 & 2.26 & 0.34\\
A3 & 31  & 1.19 & 1.21 & 0.21\\ 
\hline
\end{tabular}
\end{center}
\label{Tabla1}
\end{table}%

In Fig. \ref{fig1} we show a mosaic that summarizes the results of this first raw analysis. The spectrum of kinematic groupings is shown in the upper panel of Fig. \ref{fig1} and gives us a nice overview of the kinematic structure of the sample. 
The complete sample shows a wide RV interval, ranging more than 130 km/s. There are two coloured peaks in the spectrum corresponding to two distinct subsets of kinematic segregated bins, verifying the criterion in Eq. \ref{eq3}, peaking around 5 km/s and 75 km/s, which we will henceforth call group A and B, respectively. In addition to these segregated groups there are two other peaks with lower $\tilde{\Lambda}$ values near velocities of -5 km/s and 35 km/s that are not selected, as per Eq. \ref{eq3}. Table \ref{Tabla1} summarizes the RV characteristics of these groups. 

The lower panels of Fig. \ref{fig1} show in colour the contour levels of the star density function associated with groups A and B over the greyscale density of the whole sample. The spatial structure of the complete sample has two connected peaks near the centre in RA, and some more complex structure at lower densities. 
Group A, shown in the lower left panel of Fig. \ref{fig1}, overlaps with the main peaks in density of the whole sample, so both its spatial and radial velocity distribution is associated with the density peaks of the complete sample, and also with the typical values assigned to the NGC~ 2264 clustered population \citep[e.g.][]{Furesz06, Tobin15}. 
In contrast, group B (shown in red in the lower right panel of Fig. \ref{fig1}) is distributed mainly in the outer areas of the spatial domain of the whole sample. This means that group B members could be not associated, in both position and RV, with the central values of NGC~ 2264.

With just the data already given by the calculations of this first spectrum, we can distinguish three separated components inside group A, henceforth referred to as A1, A2 and A3. By "separated" or "unconnected" components we mean that their bins do not have any stars in common, which means that they are separated in the RV order by more than $b$ elements. The main RV characteristics of these subgroups are shown in Table \ref{Tabla1}. Subgroup A1 has the largest number of stars and the largest RV dispersion. In addition, even though they are not completely unconnected, there are wide separations between some of its bins. The spatial distribution of A1 is very similar to that of A, showing a bilobed structure with the northern peak having the largest density. Subgroups A2 and A3 have a low number of elements, low radial velocities, and their velocity distribution is also quite different from that of A1. A2 has its peak around the western-central medium density structure of the sample, around $(\Delta \rm{RA}, \Delta \rm{DEC})=(0.2,0.0)$, and A3 has its peak density in the lower lobe of the main density of the sample.

\subsection{Hierarchical analysis of group A}

It is obvious from the previous section that group A is compatible with the main population of NGC~ 2264, but there is still some structure associated with RV in this set, as indicated by the presence of unconnected components. This made us calculate a new spectrum of kinematic groupings using the set of members of the segregated bins in group A as the sample. 
In this section we will compare our analysis with the works of \citet{Furesz06} and \citet{Tobin15}, which used exploratory analysis to search for substructure in NGC ~2264.
These works use a cut in RV to restrict the sample to probable members before further analysis, but we have decided to use as probably clustered population the members of the group A described in the previous section. Our group A sample is composed of the members of the segregated bins associated to RV channels near the median RV value of the whole sample. These segregated bins are more concentrated spatially than the original raw sample with an objective statistical significance criterion (given by Eq. \ref{eq3}), as described in section 2.

\begin{figure*}
	\centering
	\includegraphics[width=\textwidth]{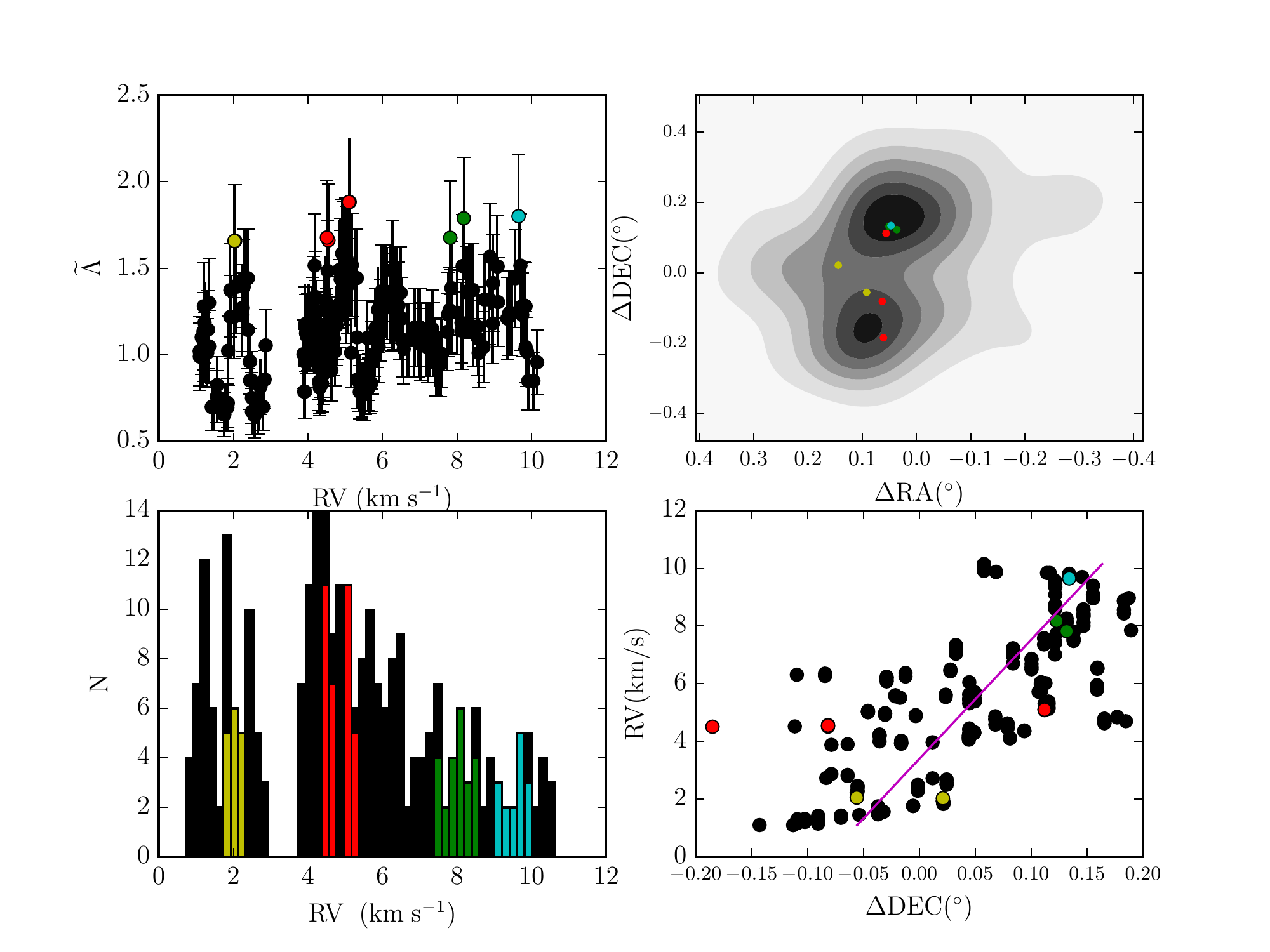}
		\caption{Four-panel summary mosaic of the analysis applied to the members group A with bin size $b=15$ and step size $s=1$. Upper left panel: Spectrum of kinematic groupings, as described in the text. The coloured circles mark the segregated velocity intervals according to the inequality defined by the equation \ref{eq3}. Group $\alpha$ (with $\widetilde{\rm{RV}}=2.04$ km/s) is shown in yellow, $\beta$ (with $\widetilde{\rm{RV}}=4.67$ km/s) is shown in red, $\gamma$ (with $\widetilde{\rm{RV}}=8.12$ km/s), and $\delta$ (with $\widetilde{\rm{RV}}=9.65$ km/s) is shown in cyan. Upper right panel: Median position of the segregated bins, over the greyscale density of the group A sample. Lower left panel: Velocity histogram of the sample, onto which the velocity histogram of the stars within the segregated intervals has been superimposed (with the same colour code as in the previous panel). Lower right panel: Median $\Delta \rm{DEC}$ vs. RV of the bins in the sample. Segregated bins are shown in red and the solid purple line is a linear fit to the data. }
		\label{fig2}
\end{figure*}

Fig. \ref{fig2} shows a summary mosaic of the results of the analysis of the 279 stars from group A with a bin size $b=15$ and step $s=1$. The upper left panel shows the SKG, and it shows four clearly segregated groups of bins, which we will name (from left to right) $\alpha$, $\beta$, $\gamma$, and $\delta$, so there is no confusion with the segregated groups described in the previous section. The median RV of $\alpha$, $\beta$, $\gamma$, and $\delta$ are 2.04 km/s, 4.67 km/s, 8.12 km/s, and 9.65 km/s, as can be seen in the RV summary shown in Table \ref{Tabla2}. There are several other peaks in the spectrum that do not fulfil Eq. \ref{eq3} but have relatively high $\tilde{\Lambda}$ values, giving the spectrum an irregular structure. This is a sign of the complex kinematical structure of this system, confirmed by the RV histogram shown in the lower left panel of Fig. \ref{fig2}, where colours represent the members of the segregated groups with the same colour code as in the spectrum. This histogram also shows two clearly separated components in group $\beta$ ($\beta1$ and $\beta2$), both with stars with an RV near 5 km/s. We have decided to show them as one same group for the sake of simplicity, given the proximity of their median RV (4.50 ad 5.11 km/s, respectively), but will detail the characteristics of $\beta1$ and $\beta2$ when needed. The majority of the stars in the sample are not in segregated bins (represented as black in the histogram of Fig. \ref{fig2}), meaning that the distribution of the associated RV channels is not significantly (with the criterion in Eq. \ref{eq3}) different from the distribution of the whole group A.

The panels on the right side of Fig. \ref{fig2} give us some insight into the spatial structure of the segregated groups. The upper right panel of Fig. \ref{fig2} shows the map of the median positions of the segregated bins over the greyscale density contours of the group A sample, using the same colour code as in the spectrum and histogram. As could also be seen in the contour plot from Fig. \ref{fig1}, group A has a bilobed north-south double peak structure. The two bins associated with segregated group $\alpha$ (in yellow), are located towards the western medium density structure and the lower density peak. When we consider the radial velocity, these bins are consistent with the unconnected components A2 and A3 previously described. The bins associated with group $\beta$ (red) are spread over the bilobed structure, the two that comprise $\beta1$ are associated with the southern and the one constituting $\beta2$ to the northern lobe. The bins associated with segregated groups $\gamma$ (green) and $\delta$ (cyan) are both clearly associated with the upper density peak. In the lower right panel of Fig. \ref{fig2} the median RV of each bin is shown against its median $\Delta \rm{DEC}$ (with segregated bins coloured with the same colour code as in previously described panels), to highlight the trend of the whole of group A: bins with higher RV tend to be located towards the north. The trend is noisy but clear, particularly when we consider a linear fit of the data, shown in Fig. \ref{fig2} as a purple line.

\begin{figure*}
	\centering
	\includegraphics[width=\textwidth]{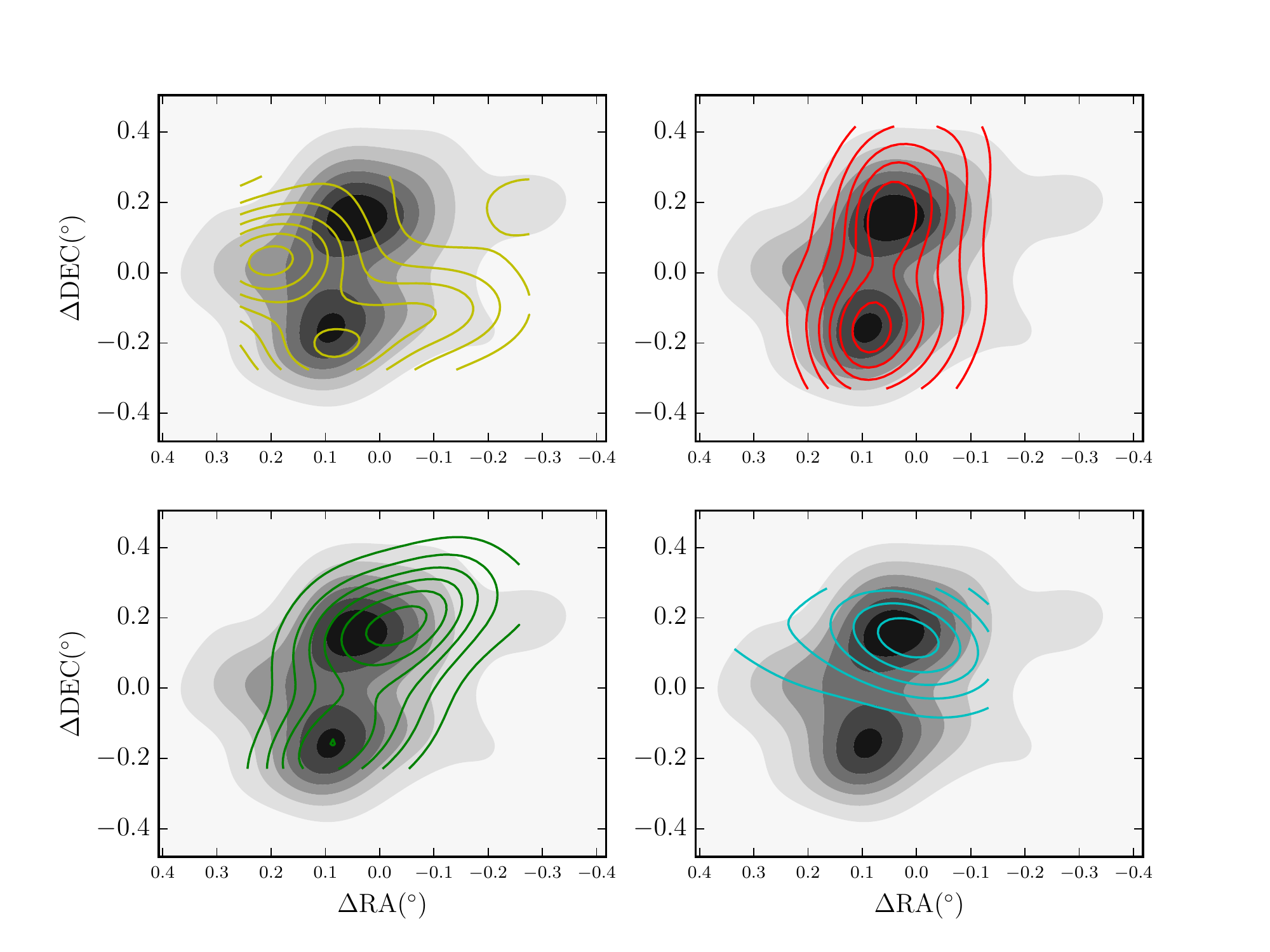}
		\caption{Spatial distribution of the four segregated groups from the analysis of group A. In each panel the coloured contours show the density star distribution of each segregated group, with the same colour code as in Fig. \ref{fig2}, over the grayscale density of the group A. From left to right and top to bottom: Group $\alpha$, in yellow, with $\widetilde{\rm{RV}}=2.04$ km/s, group $\beta$, in red, with $\widetilde{\rm{RV}}=4.67$ km/s, group $\gamma$, in green, with $\widetilde{\rm{RV}}=8.12$ km/s, and group $\delta$, in cyan, with $\widetilde{\rm{RV}}=9.65$ km/s. }
		\label{fig3}
\end{figure*}

Fig. \ref{fig3} shows in four panels the coloured contours of the density distribution of the segregated groups plotted over the greyscale density of the complete sample, following the same colour code as in Fig. \ref{fig2}: 
from left to right and top to bottom the density map of $\alpha$ (yellow), $\beta$ (red), $\gamma$ (green), and $\delta$ (cyan). Once again, the relationship of the RV of each group with the declination of its density peak is clear. Group $\alpha$ is mostly concentrated around the lower peak and the east/central structure, with the highest density in the western area. Group $\beta$, which is also the one with the most members, is distributed along the whole bilobed structure, reaching the highest density in the lower peak.  Group $\gamma$ is dense in both peaks, but mostly in the northern one, with a trend towards the east. Finally, $\delta$ group is very clearly associated with the northern peak of the bilobed structure. When we consider the unconnected components of group $\beta$ - $\beta1$ and $\beta2$ - the general RV declination trend still applies: group $\beta1$, with a lower RV, peaks in the lower lobe, while $\beta2$ peaks around the centre of the spatial domain, between both lobes.

\begin{table}
\caption{Characteristics of segregated groups $\alpha$, $\beta$, $\gamma$, and $\delta$. From left to right: Number of stars, mean radial velocity, median radial velocity and velocity dispersion}
\begin{center}
\begin{tabular}{lcccc}
\hline
Group& N & $\overline{\rm{RV}}$ (km/s) & $\widetilde{\rm{RV}}$ (km/s) & $\sigma_{\rm{RV}}$ (km/s)\\
\hline
$\alpha$&16 &2.07 &2.04 &0.15\\
$\beta$& 34& 4.82 & 4.67 &0.31 \\
$\gamma$& 23 &8.03 & 8.12&0.33 \\
$\delta$ & 15 &9.55 & 9.65 &0.28 \\ 
\hline
\end{tabular}
\end{center}
\label{Tabla2}
\end{table}%

 We will now compare these results with those obtained by \citet{Furesz06} and \citet{Tobin15}. To compare our results with \citet{Furesz06} we must first change our radial velocities from LSR to heliocentric ones. Given that the region is relatively small, we can use as a first-order approach a single correction for the velocities. Using the central coordinates of the region and using the constants given by \citet{EliasAlfCCano06} for the solar velocity, we must substract -13.24 km/s from the LSR radial velocity to obtain the heliocentric RV. This way, $\alpha$ is associated with a median heliocentric RV of 15.39 km/s, $\beta$ with 18.17 km/s, $\gamma$ with 21.47 km/s and $\delta$ with 23.00 km/s. It is also important to take into account when comparing with \citet{Furesz06} that their sample is smaller and contained in the data by \citet{Tobin15} that we analyse in this work.

The groups $\beta$, $\gamma$, and $\delta$ are in good agreement with the radial velocities shown in Fig. 5 of \citet{Furesz06}, where they plot the stars with different radial velocities between 17 and 24 km/s over the $^{13}$CO maps. In addition to coupling between stars and gas, this figure shows different substructure formed by the different velocity channels, which are in agreement with what is shown in Fig. \ref{fig3} for $\beta$, $\gamma$, and $\delta$. It is also clear from F\"uresz's Fig. 5 that structure in this velocity interval is really complex. This agrees with our spectrum in Fig. \ref{fig2}, where there are peaks at intermediate RV near 6 and 9 km/s that show a high $\tilde{\Lambda}$ value, although not significant according to our conservative criterion in Eq. \ref{eq3}. The RV gradient with DEC is also present in both works, although it is more obvious in \citet{Furesz06} because the RV domain of their sample is smaller. 

Our results are also compatible with the results given by \citet{Tobin15}. Group $\alpha$ is consistent in RV with Tobin's blueshifted population (with $|\rm{RV}|<2$ km/s). The spatial agreement is also good, as their Fig. 2 locates two clumps of blueshifted population at positions compatible with our findings: both of them near the centre of NGC~ 2264, one of them shifted towards the west and the other towards the south. 

\citet{Sungetal09}, working on the age of stars in NGC~ 2264, found a significantly larger ratio of Class I to Class II stars in the southern lobe of NGC~ 2264, whose stars would, on average, be younger than those from the northern lobe, according to \citet{SungBessell10}. Their results are compatible with the kinematical structure found in this work. Fig. 12 in \citet{Sungetal09} clearly shows a large amount of Class I objects in the southern lobe of NGC~ 2264. This younger southern lobe is compatible with the structures with RV $< 5$ km/s found with our analysis: $\alpha$ and the $\beta1$ component of group $\beta$. The older northern region would be clearly associated with our $\gamma$ and $\delta$ structures, and the component $\beta2$ would be a transition structure. In this case the spatial pattern we find in NGC~ 2264 kinematics would also be a temporal pattern.

In this section we have found the structures detected in previous works \citep{Furesz06, Tobin15} at once, using the SKG method. The spectrum of kinematic groupings shows at a single glance the RV channels more concentrated than the whole sample and thus segregated. The SKG also gives us an objective measurement of the statistical significance of the found structures. In addition, it is easily programmable and it will allow an objective and homogeneous comparison of different clusters and stellar regions.

\subsection{Discussion of Group B}

The raw and hierarchical analyses shown in previous sections give us results very compatible with those obtained by \citet{Furesz06} and \citet{Tobin15}, having the advantage of being objective analyses that can be applied in a systematic and automatic way. We have seen in the SKG in Fig. \ref{fig1} that group B has a very significant kinematic index at very high radial velocities and located in the outskirts of the spatial domain and, thus, seems associated with the field. 
This is an unexpected result, which motivates us to analyse this group B further using additional information. 

\begin{figure*}
	\centering
	\includegraphics[width=\textwidth]{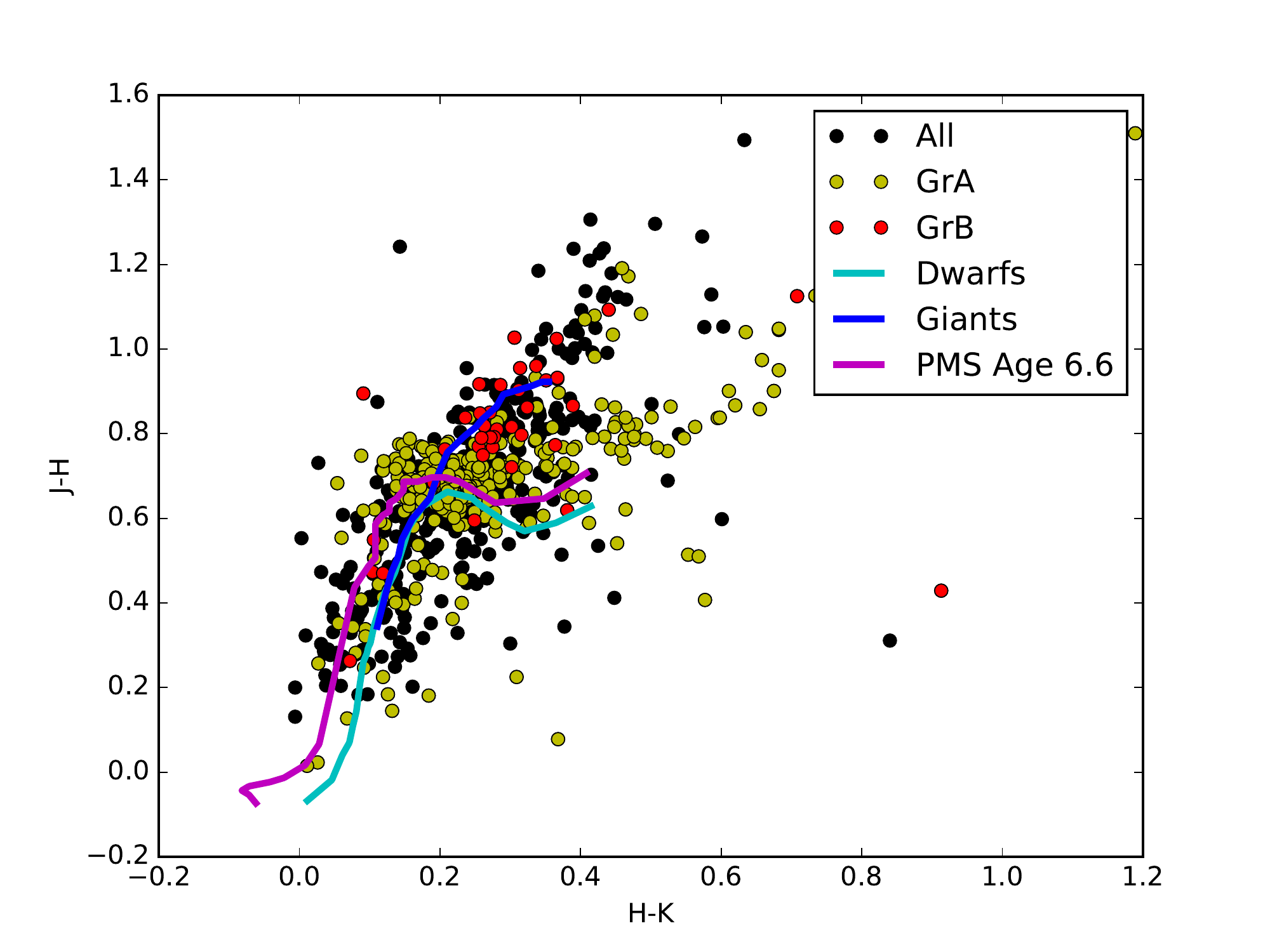}
		\caption{2MASS colour-colour diagram (J - H) vs (H - Ks)}
		\label{fig4}
\end{figure*}

Fig. \ref{fig4} shows a NIR colour-colour diagram with the colours J-H and H-K$_s$ from 2MASS data for the complete sample. The segregated members of groupings A and B are shown in colours, using the same colour code as in Fig. \ref{fig1}. We have also overplotted the intrinsic colours of dwarfs (cyan) and giants (blue) \citep[from][]{BessBrett88}, transformed to the 2MASS system using the relations in \citet{Carpenter01} with a reddening $E(B-V)=0.2$, as proposed by \citet{Sanchawala07}.  In addition, we have also overplotted an isochrone line for PMS stars with log(age)= 6.6 Myr and metallicity of 0.2 dex from \citet{Siess00}, plotted in purple after reddening correction. The stars from group A are in large proportion within the dwarf-PMS area. This is expected both from our analysis and also from the characteristics of the region, where there is intense star formation and an important PMS population. The stars from group B are grouped in the diagram, near the giant locus and well separated from the PMS stars' locus. This would be compatible with the assumption that these stars are in the background of NGC~ 2264, being underlying red-giant stars. In the following we will further discuss these hypotheses with some of the information available on the stars in group B.

If we take into account the dynamic coupling between the gas and the groups of stars shown in previous works \citep{Furesz06,Tobin15}, we can consider, as a first raw approach to a distance estimate, the kinematic distance where the RV of a star would be due to the differential rotation of the galaxy. The study by \citet{Oliver96} used this approach to characterize the molecular clouds in the Monoceros OB1 region, which contains our target area. They identified a large amount of clouds, and located them in the local arm, inter-arm and Perseus arm according to their radial velocities. Their study, however, stopped at RV=50 km/s, which is much lower than the RV of the stars in group B. 
Using the approaches and Galactic constants given by \citet{EliasAlfCCano06} we obtain, using the median coordinates and RV of group B, a distance of 6.4 kpc that would correspond to a galactocentric radius of 14.6 kpc, a distance compatible with the expected position of the Outer Arm. It is important to note that, as can be seen in Table \ref{Tabla1}, the RV dispersion in group B is large, so when we consider the minimum and maximum radial velocities in group B we get distances between 5.4 kpc to 8.9 kpc, which would lead to galactocentric radii ranging from 12.6 kpc to 17.1 kpc. Taking this into account, we may consider also that at least some of the stars in group B may belong to another structure present in this interesting quadrant of the Galaxy, the Monoceros Ring. The Monoceros Ring is a ring-like structure in the Galactic anticentre visible at a galactocentric distance of 18 kpc for $180^\circ < l < 225^\circ$ and $|b|<30^\circ$ \citep{Martin06}. Some of its characteristics are unlike those expected in the Galactic disk, so it has been suggested that it was formed by a tidally disrupted satellite galaxy, an outer spiral arm or a resonance induced by an asymmetric Galactic component \citep[see, for example,][and references therein]{Crane03}. \citet{Crane03} studied a set of presumed M giant members selected from 2MASS and obtained a distribution of galactocentric radial velocities against Galactic latitude and longitude. The radial velocities of B group members in the galactocentric system range between -37.98 and 2.50 km/s, with -26.14 km/s for the median RV. These velocities and their Galactic coordinates are in agreement with the results of \citet{Crane03} for the Monoceros Ring. The kinematic distances associated with the Galactic rotation that we have calculated are lower than 18 kpc, but the Monoceros Ring structure is not expected to be dynamically coupled to the Galactic rotation. In any case, with our current data we cannot distinguish between the Outer Arm and Monoceros Ring. 

 Both the position in the colour-colour diagram and the high radial velocities associated with group B stars seem to favour the hypothesis that these stars are at large distances, but the assumptions required to estimate distances dynamically make it hard to extract definite conclusions. For this reason, we will extend the discussion to calculating the photometric distance of the members of group B whose spectral type was found in Simbad. We have 7 of these members, but one of them is a candidate for being a white dwarf and thus will be excluded. The characteristics and distances calculated for the rest of them are shown in Table \ref{Tabla3}. The photometric distance was calculated using an average reddening of E(B-V)=0.2 and the absolute magnitudes corresponding to their spectral type assuming they are giant stars. The median photometric distance for the six selected stars is 7.0 kpc with a large spread (see Table \ref{Tabla3}), while this central value is of 6.4 kpc for the kinematic distances with a much lower dispersion (Table \ref{Tabla3}). However, there is a raw agreement between the two median crude estimates of the distance. 
 
For completion, we have also considered the SKG analysis of the complete data, lowering the confidence level required for the detection of segregation to 90\%. In this case, besides more bins associated with groups A and B, we get significant signal from a bin with median RV 41.08 km/s in the peak of the SKG centred around 35 km/s. This bin is also distributed towards the outskirts of the spatial domain and the distances associated with its RV would roughly locate its members in the Perseus arm.

 From our analysis of the segregated groups found in the field,  we have found, in this particular case, a way to roughly associate structure to another variable, outside of our initial dataset. Despite the crudeness of the estimates, we have found a way to associate the kinematic structure of RV with the 3D structure in this line of sight of the Galactic disk.

\begin{table*}
\caption{Characteristics and photometric and kinematic distances for the stars from group B with known spectral type}
\begin{center}
\begin{tabular}{cccccccc}
\hline
RA (J2000) & DEC (J2000) & RV (km/s) & V$\rm_{nomad}$ (mag)& Sp Type& Phot. Dist. Giant (kpc)& Kine. Dist. (kpc)\\
\hline
99.82367&9.55733&67.09&15.78&K0&8.6&5.8\\
100.16450&9.81111&85.01&13.61&K1&3.3&7.4\\
100.21004&9.81406&70.00&14.11&K4&5.0&6.1\\
100.40900&9.96297&73.32&14.83&K0&5.5&6.4\\
100.41946&9.5995&62.48&15.81&K1&9.1&5.4\\
100.42679&9.84206&90.15&16.67&K3&15.5&7.8\\
\hline
\end{tabular}
\end{center}
\label{Tabla3}
\end{table*}%

\section{Conclusions}

\begin{itemize}

\item With the SKG we can visualize, at a glance, the kinematic behaviour of the population in a stellar system. The algorithm is easy to implement in any pipeline aimed at analysing  the phase space of stellar systems  and has been specially designed for the study of large amounts of quality data, such as those expected from the Gaia mission. Although in this paper we have focused on the analysis of the radial velocity, the method can be easily extrapolated to other velocity components, such as the proper motions.

\item The application of the SKG method to the star-forming region NGC~ 2264 enabled us to highlight the excellent capabilities of the method for analysing the phase-space structure of a real dataset of stars. Moreover, the iterated hierarchical application of the algorithm shows the kinematic structure at different scales.

\item When we automatically calculate the SKG of the raw data we find two large significant structures (groups A and B) with very different distributions both in space and RV. Group A is clearly associated with the central positions and radial velocities of NGC~ 2264 and shows hints of internal substructure, and group B is associated with large radial velocities and spread towards the outskirts of the spatial domain. The appearance of substructure in the outer area was an unexpected result from our method.

\item The SKG allows us to distinguish structure associated with kinematic populations in star-forming regions. Depending on the scale of the sample, we can find structure associated with stellar clusters, substructure in stellar clusters or even structure in field-related velocity channels.

\item When we hierarchically iterate the analysis and study the members of group A we find at least four different subgroups separated in radial velocity, consistent with the complex kinematic structure manually found by previous works \citep{Furesz06, Tobin15}, which also found dynamical coupling of some of the structures with the surrounding gas. The results obtained by our analysis are also selfconsistent.

\item The structures found in the northern and southern lobes of NGC~ 2264 can be associated with older and younger regions, respectively, found by \citet{Sungetal09}. These resulst suggest that the kinematic structure of the phase space can be used to trace the particularities of star formation in a specific region. 

\item  The members of Group B are in the red-giant locus of the NIR colour-colour diagram of the whole sample. This may hint at B group members being giants far away in the background of NGC~ 2264. We have roughly estimated the distance to these stars with two different methods and obtained a median kinematic distance of 6.4 kpc and a median photometric distance of 7.0 kpc, both with large uncertainty. These rough distance estimates and the RV values are compatible with distant structures such as the Outer Arm or the Monoceros Ring, but we cannot distinguish between them with our current data.

\item In an area as complex as this one, the spatial structure associated with radial velocity found with our method could be roughly associated with the tridimensional structure of the Galactic disk. In this case, the SKG would work as a raw tomography of the Galactic structure in the direction of the line of sight. 

\item The structure associated with the phase-space can be used to discover populations associated with other variables not included in the analysed dataset, such as age or distance, given the intimate relationship it has with star-formation processes and the dynamical evolution of clustered stars.

\end{itemize}

\section*{Acknowledgments}

We acknowledge support from the Spanish Ministry for Economy and Competitiveness and FEDER funds through grant AYA2013-40611-P. This research has made use of the SIMBAD database and the VizieR catalogue access tool, CDS, Strasbourg, France. This work makes use of EURO-VO software, tools or services. The EURO-VO has been funded by the European Commission through contracts RI031675 (DCA) and 011892 (VO-TECH) underthe 6th Framework Program and contracts 212104 (AIDA), 261541 (VO-ICE), 312559 (CoSADIE) under the 7th Frame-work Program.

\bsp

\label{lastpage}

\end{document}